\documentclass[12pt]{article}
\usepackage{epsf}
\usepackage{latexsym}
\def\be{\begin{equation}}
\def\ee{\end{equation}}
\def\bea{\begin{eqnarray}}
\def\eea{\end{eqnarray}}

\begin{document}
\begin{titlepage}
\begin{center}
{\Large \bf William I. Fine Theoretical Physics Institute \\
University of Minnesota \\}
\end{center}
\vspace{0.2in}
\begin{flushright}
FTPI-MINN-13/45 \\
UMN-TH-3319/13 \\
December 2013 \\
\end{flushright}
\vspace{0.3in}
\begin{center}
{\Large \bf Electric discharge in vacuum by minicharged particles
\\}
\vspace{0.2in}
{\bf Xin Li$^a$  and M.B. Voloshin$^{a,b,c}$  \\ }
$^a$School of Physics and Astronomy, University of Minnesota, Minneapolis, MN 55455, USA \\
$^b$William I. Fine Theoretical Physics Institute, University of
Minnesota,\\ Minneapolis, MN 55455, USA \\
$^c$Institute of Theoretical and Experimental Physics, Moscow, 117218, Russia
\\[0.2in]

\end{center}

\vspace{0.2in}

\begin{abstract}
We discuss the possibility of a laboratory search for light minicharged particles carrying electric charge that is a small fraction $\epsilon$ of that of electron. We point out that the production of pairs of such particles in an electric field would result in a measurable discharge in vacuum of electrically charged objects. A realistic experiment may be sensitive to such particles at least down to $\epsilon \sim 10^{-8}$ if their mass is below $\sim 10^{-4}$\,eV.
\end{abstract}
\end{titlepage}

Possible existence of particles with electric charge being a small fraction $\epsilon$ of that of the electron $e$, $q_\epsilon = \epsilon e$, has been discussed theoretically and tested experimentally by probing the neutrality of matter~\cite{pk} since the very early days of particle physics. Nowadays, in spite of all the subsequent development, it is still impossible to fully exclude the existence of such minicharged `$\epsilon$ - particles' on purely theoretical grounds. Indeed, their existence can be accommodated, although in a somewhat contrived way, in GUT theories with a semisimple group~\cite{ovz}, and even in a more straighforward manner in models with an extra gauge $U(1)$ group, whose `paraphoton' has a kinetical mixing with the ordinary photon~\cite{holdom}. In the latter scheme the particles, having a `normal' charge under the extra $U(1)$ group, would appear with an electric charge with respect to the ordinary photon being proportional to the mixing parameter. Thus the $\epsilon$-particles can be sought for in a vast range of their fractional charge $\epsilon$ and their mass $m_\epsilon$. In particular such particles were recently discussed in connection with the bounds on the stability of the photon~\cite{heeck} , and also on non standard properties of neutrinos~\cite{studenikin}. The updated version~\cite{mm} of the experimental probe of the electric neutrality of matter and the charge conservation in $\beta$ decay place a very strong bound on the charge of the neutrinos~\cite{ovz,fjlv}, $\epsilon_\nu < 10^{-21}$, but for new particles the value of $\epsilon$ can be substantially larger. A comprehensive review of the limits on the parameters ($\epsilon, \, m_\epsilon$) from laboratory experiments and astrophysical considerations can be found in Ref.~\cite{dhr}, with the strongest upper limit on $\epsilon$ being in the range $10^{-13} - 10^{-14}$ and arising from considering the cooling of Red Giants and White Dwarfs, so that this limit applies if the mass $m_\epsilon$ does not exceed few keV. The bounds from laboratory experiments are considerably weaker. The best upper bound $\sim 10^{-6}$ on $\epsilon$ at the mass $m_\epsilon$ below $\sim 10^{-2}$\,eV comes from the estimate~\cite{gjr} of the energy loss due to production of pairs of minicharged particles by high $Q$-factor RF cavities used in accelerators. It was subsequently noticed~\cite{jaeckel} that for very small $m_\epsilon$, below $10^{-7}$\,eV, an upper limit approximately $\epsilon < 5 \times 10^{-7}$ can be inferred from the tests of the Coulomb law~\cite{cl1,cl2,cl3}, that would be modified~\cite{uehling} by the vacuum polarization due to the minicharged particles at the typical distances of about 1\,m used in these tests.

In this paper we consider another effect that may allow a laboratory test for minicharged particles with mass up to approximately $10^{-4}$\,eV and $\epsilon$ down to $\sim 10^{-8}$, namely the discharge of electrically charged bodies in vacuum due to the Schwinger~\cite{sauter,schwinger} process of pair creation in an electric field. The pair production of minicharged particles was mentioned in Ref.~\cite{jaeckel2} and subsequently used as the mechanism for the energy loss in RF cavities in Ref.~\cite{gjr}. According to the estimates that will be presented below, an experiment of the type that we discuss here, based on the same Schwinger process, is likely more straightforward and has a better sensitivity to light minicharged particles. This hypothetical experiment involves a sphere of radius $R$ initially charged to a potential $V_0$ and placed in vacuum. The electric field near its surface creates pairs of minicharged particles. Assuming for definiteness that the sphere is carrying a positive charge, the created positive $\epsilon$-particles will be expelled (to infinity) by the Coulomb field, while the negative ones will stay near the sphere and screen its charge. As the pairs are constantly produced, the effective charge of the sphere should decrease with time, which can be measured e.g. by the variation of the attractive force between the charged sphere and a distant grounded conductive plane. We estimate the characteristic time of the discharge for a sufficiently small sphere and small mass $m_\epsilon$ as 
\be
t_d = \left ( \epsilon^3 \, {\alpha \over \pi^2} \, e \, V_0 \right )^{-1} \approx 100\,{\rm days} \, \left ( {10^{-8} \over \epsilon} \right )^3 \, \left ( {100\,{\rm kV} \over V_0} \right )~.
\label{td}
\ee
In other words, the sphere will lose approximately 1\% of its charge in one day, if $\epsilon \approx 10^{-8}$ and $V_0 \approx 100$\,kV. The applicability of the expression (\ref{td}) also assumes that the radius $R$ of the charged sphere, satisfies certain conditions, which will be discussed further in the text. In our `benchmark' estimates we assume $V_0 \approx 100$\,kV and $R=1$\,cm, which are hopefully not very ambitious assumptions about what is achievable in a laboratory experiment. Also we use the Gauss' system of electrostatic units throughout this paper, so that $e^2 = \alpha \approx 1/137$.

The number $N$ of pairs created in a uniform electric field $E$ per unit time and per unit volume for particles with charge $q$ and mass $m$ is given by~\cite{nikishov}
\be
N_f = {(q E)^2 \over 4 \pi^3} \, \exp \left ( - {\pi m^2 \over q E} \right )~~~{\rm and}~~~ N_s = {(q E)^2 \over 8 \pi^3} \, \exp \left ( - {\pi m^2 \over q E} \right )
\label{npair}
\ee
for respectively spin 1/2 fermions and scalars. It can be noted that, as first pointed out in Ref.~\cite{nikishov} and more recently discussed in Ref.~\cite{cmg}, this quantity is different from the rate $\Gamma$ of decay of the vacuum (i.e. of the zero-pair state) in the same electric field, which is given~\cite{schwinger}  by the respective expressions
\be
\Gamma_f = {(q E)^2 \over 4 \pi^3} \, \sum_{n=1}^{\infty} \, {1 \over n^2} \, \exp \left ( - {\pi m^2 \over q E} \right )~~~{\rm and}~~~ \Gamma_s = {(q E)^2 \over 8 \pi^3} \, \sum_{n=1}^{\infty} \, {(-1)^{n+1} \over n^2} \, \exp \left ( - {\pi m^2 \over q E} \right )~.
\label{gpair}
\ee
(In other words, the rate for pair creation $N$ is given in each case by only the first term in the sum for $\Gamma$.) Clearly, it is the rate for pair creation (\ref{npair}) that is of relevance for the problem of the electric discharge which is discussed here. 

It is also evident from Eq.(\ref{npair}) that the mass of the particles does not suppress their production if $m^2 \ll qE$, and this is the limit that we will assume, given that at larger mass the tunneling exponent effectively cuts off the production rate. The limits on $m_\epsilon$ and $\epsilon$ that can be achieved by an experiment of the discussed type thus critically depend on the achievable strength of the field $E$ and its distance scale $R$. 
The production of $\epsilon$ particles near the surface of the sphere is not suppressed by the exponent if their mass satisfies the condition
\be
m_\epsilon^2 \ll {\epsilon \, e \, V_0 \over R} \sim (10^{-4}\, {\rm eV})^2 \,  \left ( {\epsilon \over 10^{-8}} \right ) \, \left ( {V_0 \over 100\,{\rm kV}} \right ) \, \left ( {1\,{\rm cm} \over R} \right )~.
\label{mcond}
\ee
The particles are then produced at the characteristic distance scale determined by the field strength
\be
\ell \sim (\epsilon \, e V_0/R)^{-1/2} \sim 0.1\,{\rm cm} \, \left ( {\epsilon \over 10^{-8}} \right )^{-1/2} \, \left ( {V_0 \over 100\,{\rm kV}} \right )^{-1/2} \, \left ( { R \over 1\,{\rm cm} } \right )^{1/2},
\label{ellch}
\ee
so that one can neglect the non uniformity of the field and use the expressions (\ref{npair}) for a constant field as long as $\ell \ll R$, which requirement can be written as a condition for $\epsilon$:
\be
\epsilon \gg (e \, V_0 \, R)^{-1} \sim 10^{-10} \, \left ( {100\,{\rm kV}} \over  V_0  \right ) \, \left ( {1\,{\rm cm} \over R} \right )~.
\label{conde}
\ee

If the interaction of the $\epsilon$-particles with the matter of the sphere is only the electromagnetic interaction due 
to their minicharge, they essentially pass freely through the test sphere and the negative ones are captured by the Coulomb field at the maximal distance determined by the distance at which they were produced, and which is of order $R$. In the case where the produced particles do have other interaction with the ordinary matter they are captured in the material of the sphere and thus also stay within the radius $R$. The total rate of accumulation of the negative charge  from pairs produced in the volume around the sphere is given by integrating the expression (\ref{npair}) over the volume occupied by the Coulomb field outside the sphere. For the fermion case one readily finds (for scalar $\epsilon$-particles the numerical coefficient is twice smaller)
\be
{d Q \over dt} = - \epsilon^3 \, {\alpha \over \pi^2} \, e \, {Q_0^2 \over R} = - \epsilon^3 \, {\alpha \over \pi^2} \, Q_0 \, e \, V_0~,
\label{dqdt}
\ee
where $Q_0 = V_0 \, R$ is the initial electric charge on the sphere. Thus the characteristic time of initial discharge $t_d = Q_0/(dQ/dt)$ is estimated as given by Eq.(\ref{td}). At later times one should take into account the back reaction of the screening by the accumulated negative charge on the Coulomb field and thus on the rate of the charge accumulation. The specific calculation then depends on the spatial distribution of the screening charge which in its turn depends on the interaction of the $\epsilon$-particles with the matter, and also on their mass. The latter dependence arises from the fact that the captured negative particles radiate due to their acceleration in the Coulomb field. Indeed, the total power radiated by a (generally relativistic) particle in an electric field is given by the well known formula [see e.g. in the textbook \cite{ll}, Eq.(73.7) ]
\be
{dW \over dt} = {2 \, \epsilon^4  \, e^4 \, E^2 \over 3 \, m_\epsilon^2}
\label{rp}
\ee
with $W$ being the radiated energy. Since the initial energy of the created particle is of order $W_0=\epsilon \, e \, V_0$, and $E \sim V_0/R$, one estimates the characteristic time over which the $\epsilon$-particles lose their energy by radiation as
\be 
t_r = {W_0 \over dW/dt} \sim   {m_\epsilon^2 \, R^2 \over \epsilon^3 \, \alpha \, e \, V_0}~, 
\label{tr}
\ee
so that the radiation time can be compared with the discharge time in Eq.(\ref{td}) as
\be
t_r \sim t_d \, m_\epsilon^2 \, R^2~.
\label{trtd}
\ee
Since we make no assumptions here about the value of the product $m_\epsilon R$, as long as the conditions (\ref{mcond}) - (\ref{conde}) are satisfied, the present treatment allows for arbitrary relative values of the characteristic time for radiation and discharge. We believe that at this point it would be premature to analyze in detail the process of discharge at later times, when the effect is of order one, since it is plausible that even a small (say, 1\%) initial discharge would be quite detectable in a realistic experiment.

It appears helpful, for understanding the effects of light minicharged particles on the Coulomb field of charged bodies, to discuss the relation between the modification of the field due to the vacuum polarization and the production of pairs of the $\epsilon$-particles. If $m_\epsilon$ is small and satisfies the condition (\ref{mcond}) the vacuum effects are not determined by the mass, but rather by the largest of the scales set by the `momentum transfer' $p \sim 1/r \sim 1/R$ and the field scale $(\epsilon \, e \, E)^{1/2} \sim (\epsilon \, e \, V_0/R)^{1/2}$, which is the typical momentum supplied by the field to the particle. In a situation where the former momentum scale is larger, the vacuum polarization effects can be calculated from a standard loop with the $\epsilon$-particles and can be described by the effective Lagrangian in a purely electric field (see e.g. in the textbook \cite{blp})~\footnote{For definiteness the formulas for spin 1/2 particles are discussed. For scalars the coefficient in front  of the logarithm is four times smaller.} 
\be
L_{eff} = - {\epsilon^2 \, \alpha \over 24 \pi^2} \, E^2 \, \log (r^2 \Lambda^2)~,
\label{smalle}
\ee
where the precise meaning of the parameter $\Lambda$ depends on the chosen convention for renormalization of $\alpha$ and is of no significance for the present discussion. The variation of the effective Lagrangian (\ref{smalle}) gives rise to the modification~\cite{uehling} of the Coulomb potential, discussed~\cite{jaeckel} in connection with the bounds on the $\epsilon$-particles. If, however, the electric field is strong, $\epsilon \, e \, E \gg 1/r$, then the vacuum polarization loop should be calculated with the exact propagator in the field rather than with a free-particle Green function. (It should be noted that in this case the calculation of the vacuum polarization is affected even inside a charged sphere, where locally the unmodified Coulomb field strength is zero. This is due to the fact that the propagators in the loop also include particle paths that extend outside the sphere.) In this limit the effective Lagrangian, as determined by the well known Euler-Heisenberg calculation~\cite{blp}, has the form
\be
L^{'}_{eff}= {\epsilon^2 \, \alpha \over 24 \pi^2} \, E^2 \, \log \left ( {i \, \epsilon \, e \, E \over \Lambda^2 } \right )~.
\label{largee}
\ee
The real part of this expression describes the modification to the Coulomb field, while the imaginary part describes the decay of the vacuum state: 
\be
\Gamma_f = 2 \, {\rm Im} L^{'}_{eff} = {\epsilon^2 \, \alpha \over 24 \pi} \, E^2~,
\label{gfeff}
\ee
which is the $m \to 0$ limit of the first expression in Eq.(\ref{gpair}).

Clearly, in a spherical geometry  the condition for the field scale to be dominant, and the effective Largangian (\ref{largee}) to be appropriate, is exactly the inequality (\ref{conde}). One can then notice that for the typical parameters of the most sensitive experiment~\cite{cl3}, $V_0 = (40 \div 70)$\,kV and $R \sim 1$\,m, this condition is satisfied as long as $\epsilon$ is larger than $\sim 10^{-11}$ and is thus well applicable at the claimed~\cite{jaeckel} bound $5 \times 10^{-7}$ for $\epsilon$. Under these circumstances the applicability of the Uehling correction derived from the effective Lagrangian (\ref{smalle}) is not justified, and the interpretation of the experiment~\cite{cl3} in terms of a bound for $\epsilon$ has to be modified, although it is not clear at present how strong such modification would be numerically. It can be also noted that the discussed here effect of accumulation of the $\epsilon$ particles due to the pair production [related to the imaginary part of the Lagrangian (\ref{largee})] is proportional to time, and can thus be much greater than that of the real part of the vacuum polarization. For this reason we believe that a laboratory measurement of the discussed here type can achieve a higher sensitivity to minicharged particles.

In summary. We estimate that the production of pairs of light minicharged particles by electric field results in a realistically measurable discharge of charged objects in vacuum for the value of $\epsilon$ as low as $\sim 10^{-8}$. At such $\epsilon$ a sphere of the radius 1\,cm charged to the potential 100\,kV would lose about 1\% of its charge after one day, provided that the mass of the hypothetical particles is less than $\sim 10^{-4}\,$eV. We also find that for objects at a high, but still realistic, potential the discharge through the pair creation is the more readily observable than the modification of the Coulomb's law due to the vacuum polarization.

This work is supported, in part, by the DOE grant DE-FG02-94ER40823.

\end{document}